\documentclass[11pt,a4paper]{article}
\usepackage{jcappub}
\usepackage{aasmacros}

\usepackage{rotating} 
\usepackage{xspace}
\usepackage{hyperref}
\usepackage{soul}
\usepackage[dvipsnames]{xcolor}
\usepackage[normalem]{ulem}
\usepackage{lineno}

\usepackage{bbold}
\usepackage{amssymb,amsfonts,amsthm}
\usepackage{epsfig,graphicx}
\usepackage{url} 
\usepackage{bm} 
\usepackage{slashed}
\usepackage{color}
\usepackage{pstricks}
\usepackage{fancyhdr}
\usepackage{textcomp}
\newcommand{\unit}{\leavevmode\hbox{\small1\kern-3.6pt\normalsize1}}

\def\mdm{m_{\chi}}

\def\locald{\rho_0}

\def\mnucl{m_{N}}
\def\redT{\mu_N}

\newcommand{\Emax}{E_{\textrm{R}}^{\textrm{max}}}
\newcommand{\Emin}{E_{\textrm{R}}^{\textrm{min}}}

\newcommand{\mwimp}{m_\chi}

\def\op#1{{\cal O}_{#1}}

\title{Opening the energy window on direct dark matter detection}

\author[a]{Nassim Bozorgnia,}
\author[a]{David G.~Cerde\~no,}
\author[a]{Andrew Cheek,}
\author[b]{and Bjoern Penning}

\affiliation[a]{Institute for Particle Physics Phenomenology, Department of Physics,\\
Durham University, Durham DH1 3LE, UK} 
\affiliation[b]{Department of Physics, Brandeis University, 415 South Street,\\
Waltham, MA 02453, USA}

\abstract{
In this article we investigate the benefits of increasing the maximum nuclear recoil energy analysed in dark matter (DM) direct detection experiments. We focus on elastic DM-nucleus interactions, and work within the framework of effective field theory (EFT) to describe the scattering cross section. In agreement with previous literature, we show that an increased maximum energy leads to more stringent upper bounds on the DM-nucleus cross section for the EFT operators, especially those with an explicit momentum dependence. In this article we extend the energy region of interest (ROI) to show that the optimal values of the maximum energy for xenon and argon are of the order of $500$~keV and $300$~keV, respectively. We then show how, if a signal compatible with DM is observed, an enlarged energy ROI leads to a better measurement of the DM mass and couplings. In particular, for a xenon detector, DM masses of the order of 200 GeV (2~TeV) or lower can be reconstructed for momentum-independent (-dependent) operators. We also investigate three-dimensional parameter reconstruction and apply it to the specific case of scalar DM and anapole DM. We find that opening the energy ROI is an excellent way to identify the linear combination of momentum-dependent and momentum-independent operators, and it is crucial to correctly distinguish these models. Finally, we show how an enlarged energy ROI also allows us to test astrophysical parameters of the DM halo, such as the DM escape speed. 
}

\arxivnumber{IPPP/18/92}

\begin{document}
\maketitle


\section{Introduction}
\label{sec:introduction}
\setcounter{equation}{0}

Over the past decades, direct detection experiments have searched for DM particles through their elastic scattering off a target material in underground detectors \cite{Goodman:1984dc, Primack:1988zm}. A worldwide experimental effort has resulted in extremely sensitive experiments, which have probed DM interactions with ordinary matter with unprecedented precision, albeit not observing any confirmed excess. This has resulted in stringent upper bounds on the DM elastic scattering cross section with nuclei \cite{Aprile:2018dbl, Akerib:2017kat, Cui:2017nnn, Agnes:2018fwg,Agnese:2018gze,Petricca:2017zdp,Aguilar-Arevalo:2016ndq} (bounds on DM-electron interaction and inelastic scattering can also be derived). Currently, liquid xenon experiments, such as LUX \cite{Akerib:2016vxi}, XENON1T \cite{Aprile:2018dbl}, and PandaX \cite{Cui:2017nnn}, dominate the search for DM masses above 10~GeV, whereas the low-mass region of $\sim 10$~GeV and less is being explored by a  range of experiments using noble liquids and crystals as targets. The next-generation of liquid noble gas detectors, using xenon (such as LZ \cite{Dobson:2018xxl}, XENONnT \cite{aprile:2015xxl}, and DARWIN \cite{Aalbers:2016jon}) or argon (DarkSide-20k \cite{Aalseth:2017fik} and DEAP \cite{Amaudruz:2017ekt}) will probe the parameter space of weakly interacting massive particles (WIMPs) with unprecedented precision.

In order to optimise the discovery potential of current and future detectors, one must have an excellent control over the experimental background, either by reducing it through the use of  shielding and  employing extremely radiopure materials or by understanding any source of irreducible background. Likewise, the characteristics of the expected signature from DM interactions must be well understood, as this defines the region of interest in which a signal might be expected. In direct detection experiments, DM particles can undergo elastic collision with the target nuclei, leaving an energy deposit (spectrum) in the keV--MeV range.

The most general DM-nucleus scattering cross section can be expressed in terms of non-relativistic EFT operators \cite{Fan:2010gt,Fitzpatrick:2012ix,Anand:2013yka,Dent:2015zpa,Catena:2014epa}, which originate from different terms in the microscopical Lagrangian that describes DM-quark interactions. Traditionally, the velocity and momentum-independent spin-independent (SI) and spin-dependent (SD)  operators, are considered to interpret results from direct detection experiments and derive bounds on the DM-nucleon scattering cross section. For these operators, the expected nuclear recoil spectrum is approximately exponential, with most of the signal concentrated in the keV and sub-keV region and a slope that decreases with increasing DM mass. Thus, in order to capture most of the DM signal, a great effort is made in lowering the experimental energy threshold. This also helps probing lighter DM particles, which leave a smaller energy deposit in the detector.

However, the shape of the DM spectrum changes substantially for other EFT operators, mainly for those with a non-trivial momentum dependence, which display a characteristic bump at large recoil energies \cite{Schneck:2015eqa}. Likewise, the higher end of the recoil spectrum is particularly sensitive to some of the astrophysical parameters that describe the DM velocity distribution in the Solar neighbourhood, for example, to changes in the DM escape speed from the Galaxy. All of this motivates widening the energy window analysed in direct detection experiments.

The possibility of extending the energy range of direct detection experiments to include the high energy end of the recoil spectrum has been addressed in the literature.  Earlier work in this area \cite{TuckerSmith:2001hy,Graham:2010ca,Dienes:2014via,Bramante:2016rdh} studied inelastic and exothermic DM~\cite{Bramante:2016rdh, Barello:2014uda, Graham:2010ca}, and more recent works have pointed out the effects on EFT operators for elastic scattering \cite{Gluscevic:2015sqa,Aprile:2017aas,Gelmini:2018ogy}. The prospects of reconstructing the DM speed distribution and particle physics parameters from direct detection data for an extended energy window up to 1 MeV has also been studied in ref.~\cite{Peter:2011eu} for the canonical SI cross section. In this article we will study in detail the advantages of extending the energy window in direct detection experiments, with emphasis on xenon and argon based detectors. We consider large values for the maximum nuclear recoil energy, and study the optimal energy ranges for xenon and argon to maximise the sensitivity to EFT operators. Then, assuming a DM detection, we investigate how an extended energy range affects the reconstruction of DM parameters (mass and couplings), and we take into account the effect of nuclear form factors and the neutrino floor in this energy range. We also extend our study to investigate how a larger energy range can give us access to some astrophysical parameters of the DM halo.

This article is organised as follows. In section~\ref{sec:direct}, we review the principles of direct DM detection and the computation of the expected rate of DM events. In section~\ref{sec:benefits}, we explain the various aspects of DM detection that would benefit from an increase in the maximum energy, thereby utilizing a larger part of the overall spectrum. This will lead to a significant improvement in the exclusion limits for momentum-dependent EFT operators in section~\ref{sec:limits}, a better reconstruction of the DM mass and couplings in section~\ref{sec:reconstruction}, and a better sensitivity to astrophysical parameters such as the escape speed in section~\ref{sec:astro}. Finally, our conclusions are presented in section~\ref{sec:Conclusion}.

\section{Principles of dark matter scattering and signal generation}
\label{sec:direct}

The expected event rate from the elastic scattering of a DM particle, ${\chi}$, with mass $\mwimp$ off a target nucleus with mass $\mnucl$ is given by 
\begin{equation}
N =\frac{\locald \, \epsilon}{\mnucl\,\mwimp}\int_{\Emin}^{\Emax}\varepsilon(E_R)\, dE_R \int_{v_{min}} d {\bf v}\, v f({\bf v}, t)\, \frac{d\sigma_{{\chi}N}}{dE_R}  \,,\label{eq:drate}
\end{equation}
where $d\sigma_{\chi N}/dE_R$ is the DM-nucleus scattering cross-section, $\locald$ is the local DM density, ${\bf v}$ is the relative velocity between the DM particle and the nucleus, $f({\bf v}, t)$ is the DM velocity distribution in the detector frame normalized to unity, and $\epsilon$ is the total exposure, given by the product of the  detector mass 
and the run-time. The total DM count is calculated by integrating the differential event rate over the nuclear recoil energy, $E_R$, within an energy ROI, defined as the range between the minimum (threshold) energy, $\Emin$, and a maximum energy $\Emax$. The energy ROI is normally chosen to maximise the potential for DM discovery while keeping the background under control. The detector efficiency $\varepsilon(E_R)$ is an energy-dependent function that describes the fraction of recoil events that will be observed in the given energy ROI. For simplicity, we will assume $\varepsilon(E_R)=1$.

The integration over the DM velocity is performed from the minimum DM speed needed to induce a nuclear recoil of energy $E_{R}$, $v_{min}=\sqrt{\mnucl E_R/2\redT^2}$, where $\redT$ is the DM-nucleus reduced mass. It is customary to assume the Standard Halo Model (SHM)~\cite{Drukier:1986tm} for the DM distribution in our Galaxy. The local DM velocity distribution in the SHM is described in terms of an isotropic Maxwell-Boltzmann distribution in the Galactic rest frame, truncated at the escape speed from the Galaxy. However, the true DM distribution may be significantly different from the SHM. In the rest of this work up to section \ref{sec:astro} we will assume the SHM for the local DM distribution with parameters given in table~\ref{tab:astroparam} (in section \ref{sec:astro}). We will consider an alternative halo model motivated by recent hydrodynamic simulations in section \ref{sec:astro}.

The DM-nucleus differential cross section, $d\sigma_{{\chi}N}/dE_R$, is traditionally split into an SI and SD contribution. However, the general Lagrangian describing DM interactions with nuclei in the non-relativistic limit is much more diverse and can be described in terms of an EFT which features different operators, some of which display a non-trivial dependence on the DM velocity and the momentum exchange \cite{Fan:2010gt,Fitzpatrick:2012ix},
\begin{equation}
{\cal L}_{\text{int}}=\sum_{\tau} \sum_i c_i^{\tau}\mathcal{O}_i \overline{\chi} \chi \overline{\tau} \tau\ .
\label{eq:eft}
\end{equation}
The EFT operators can be understood as the non-relativistic reduction from a microscopic Lagrangian that describes the DM interactions with quarks \cite{Dent:2015zpa}. In this expression, $\tau$ can either represent proton and neutron interactions or isoscalar and isovector interactions, and the list of viable operators can be found in table~\ref{tab:eft}~\cite{Fitzpatrick:2012ix}, classified according to their dependence on the momentum exchange, $q$. We have limited ourselves to DM candidates of spin $1/2$ or $1$, but results for higher spin can be found in ref.~\cite{Dent:2015zpa}.
\begin{table}[!t]
\begin{center}
\begin{tabular}{c|c|c}
\hline
{\small $q$-independent}& {\small $q$-dependent} &{\small $q^2-$dependent}\\
\hline
{\small 
\begin{minipage}{3.5cm}
\ \\
$\mathcal O_1 = 1_{\chi} 1_N$\\
$\mathcal O_4 = {\bf \hat S}_{\chi} \cdot {\bf \hat S}_N$ \\
$\mathcal O_7 = {\bf \hat S}_N \cdot {\bf \hat v}^{\bot}$ \\
$\mathcal O_8 = {\bf \hat S}_{\chi} \cdot {\bf \hat v}^{\bot}$ \\
$\mathcal O_{12} = {\bf \hat S}_{\chi} \cdot \left [ {\bf \hat S}_N \times {\bf \hat v}^{\bot} \right ]$ \\
\end{minipage}
}
&
{\small 
\begin{minipage}{6cm}
\ \\
$\mathcal O_3 = i {\bf \hat S}_N \cdot \left [ \frac{ {\bf \hat q}}{m_N} \times {\bf \hat v}^{\bot} \right] $\\
$\mathcal O_5 = i {\bf \hat S}_{\chi} \cdot \left [ \frac{ {\bf \hat q}}{m_N} \times {\bf \hat v}^{\bot} \right]$ \\
$\mathcal O_9 = i {\bf \hat S}_{\chi} \cdot \left [ {\bf  \hat S}_N \times  \frac{ {\bf \hat q}}{m_N} \right ]$ \\
$\mathcal O_{10} = i {\bf \hat S}_N \cdot \frac{ {\bf \hat q}}{m_N}$ \\
$\mathcal O_{11} = i {\bf \hat S}_{\chi} \cdot \frac{ {\bf \hat q}}{m_N}$ \\
$\mathcal O_{13} = i \left [ {\bf \hat S}_{\chi} \cdot {\bf \hat v}^{\bot}  \right ]  \left [ {\bf \hat S}_N \cdot  \frac{ {\bf \hat q}}{m_N} \right ]$ \\
$\mathcal O_{14} = i \left [ {\bf S}_{\chi} \cdot  \frac{ {\bf q}}{m_N} \right ]  \left [ {\bf \hat S}_N \cdot  {\bf \hat v}^{\bot} \right ]$ \\
$\mathcal O_{15} = - \left [ {\bf\hat S}_{\chi} \cdot  \frac{ {\bf \hat q}}{m_N} \right ] \left [\left ( {\bf  \hat S}_N \times  {\bf \hat v}^{\bot} \right ) \cdot \frac{ {\bf  \hat q}}{m_N} \right ]$ 
\end{minipage}
}
&
{\small 
\begin{minipage}{4.2cm}
\ \\$\mathcal O_6 = \left [ {\bf \hat S}_{\chi} \cdot  \frac{ {\bf \hat q}}{m_N} \right ]  \left [{\bf\hat  S}_N \cdot  \frac{ {\bf\hat  q}}{m_N} \right ]$ \\
\end{minipage}
}
\\
\ &&\\
\hline
\end{tabular}
\caption{List of non-relativistic EFT operators for spin-$1/2$ and spin-$1$ DM particles, classified according to their dependence on the momentum exchange. }
\label{tab:eft}
\end{center}
\end{table}

The total DM-nucleus cross section is calculated by adding these contributions coherently, using nuclear wave functions, which results in the following expression,
\begin{equation}
\frac{d\sigma_{{\chi}N}}{dE_R}=\frac{m_N}{2\pi m_v^4} 
\frac{1}{v}\sum_{ij}\sum_{\tau, \tau'=0,1} c^{\tau}_ic^{\tau'}_j\mathcal{F}^{\,\tau,\tau'}_{i,j}(v^2,q^2)
\ .
\label{eq:dsigma}
\end{equation}
Here, $\mathcal{F}_{i,j}^{\,\tau,\tau^\prime}$ are the nuclear form factors (see e.g., Refs.\,\cite{Fitzpatrick:2012ix,Anand:2013yka} for their expressions in the isospin and nuclear basis, respectively).  There are interference terms between the two isospin-components within each operator, as well as between the following pairs of EFT operators, ($\op{1},\op{3}$), ($\op{4},\op{5}$), ($\op{4},\op{6}$), and ($\op{8},\op{9}$). We choose the couplings $c_i$ to be dimensionless, using the Higgs vacuum expectation value, $m_v=264$~GeV,  as a normalisation factor, following the prescription of refs.~\cite{Anand:2013yka,Catena:2014epa}.

As an explicit example of a particle DM model that features momentum-dependent interactions, we will consider the case of anapole DM, defined by the interaction Lagrangian $\mathcal{L}_{\textrm{int}}=\mathcal{A}\,\overline{\chi}\gamma^{\mu}\gamma^5\chi\partial_{\nu} F^{\mu\nu}$. This is the only dimension six operator that interacts with the electromagnetic field for Majorana particles \cite{Ho:2012bg}. In the non-relativistic limit, the effective operator for anapole interactions, $\op{\cal A}$, is a linear combination of the momentum-independent operator $\op{8}$ and the momentum-dependent $\op{9}$ with the Lagrangian as follows\footnote{The relative sign between the two terms depends on the convention used for the choice of the transfer momentum direction.} \cite{Gresham:2014vja,DelNobile:2018dfg,Kavanagh:2018xeh},
\begin{equation}
\mathcal{L}_{\cal A}=\sum_{N=p,n}\mathcal{A}\, e(2 Q_N \op{8} - g_N\op{9})\overline{\chi}\chi\overline{N}N,
\end{equation}
where $e$ is the electron charge, $Q_N$ is the nucleon charge, and $g_N$ are the nucleon g-factors ($g_p=5.59$ and $g_n=3.83$). We can parameterize the coupling strength as $\sigma_{\mathcal{A}} = \mathcal{A}^2 \mu_{N}^2/\pi$ \cite{Gresham:2014vja}.

In this work, we focus on the effect of higher recoil energies in noble liquid detectors which employ either liquid xenon or liquid argon. Although the qualitative results can be extrapolated to other targets and experiments, the main advantage is for heavy DM particles, where noble liquid detectors excel. Both detectors are dual phase time projection chambers (TPC), utilizing two type of signals: a prompt photon signal from the scintillation in the liquid xenon and a proportional charge signal amplified in the gas phase. The ratio of the two allows to distinguish electron from neutron recoils. The position of the interaction in the TPC can be determined from the drift times and light pattern of the signals, allowing to define a background-free fiducial volume due to the excellent self-shielding of xenon.

The primary DM signature is the spectrum of nuclear recoils reconstructed using the resulting charge and light signals in liquid and double phase noble gas detectors. There is also a series of background sources that limit the experimental sensitivity. These originate either from natural radioactivity, mostly from  naturally occurring $^{238}$U and $^{232}$Th chains, as well as cosmic muon and spallation induced fission products.  The dominant backgrounds for many DM searches are neutrons that interact with nuclei in the detector target via elastic scattering. This produces a nuclear recoil similar to the expected signal. High energy cosmogenic neutrons of up to a few GeV might be produced by spallation reactions of cosmic muons on nuclei in the detector or the surrounding rock. Further $(\alpha, n)$ reactions where an $\alpha$ particle can initiate nuclear reactions in the target nucleus while emitting a neutron and  spontaneous fission reactions produce neutrons at moderately low energies of around a few MeV. Just as for standard low recoil DM searches, such energy depositions can end up in the region of interest for our searches. Typically the background levels decrease by orders of magnitude from the low energy DM search region of $10~\rm{keV}$ to higher energies above $100~\rm{keV}$~\cite{Mei:2005gm}. Thus searches in the high nuclear recoil energy range can achieve very good sensitivities.

DM is expected to scatter only once in the detector because of its low interaction probability. In contrast most backgrounds are expected to scatter multiple times. Therefore experimentally these backgrounds are identified and rejected by removing multiple scatter events. In the case of high energy nuclear recoils certain instrumental background processes (such as accidental coincidence between single-electron and single-photon noise) might become relatively more important. Another type of instrumental background relevant for the high nuclear energy regime occurs when one of the multiple scatters takes place outside of the sensitive volume and the other one inside the sensitive volume. Then the multiple scatter is mis-characterized as single-scatter and also its charge yield is not properly reconstructed. These effects are highly dependent on detector geometry and cannot be generally assessed here.

When reconstructing the energy of an incoming particle from the measured light and charge yields, calibration data is necessary. Current noble liquid detectors have developed a comprehensive understanding of backgrounds and calibrations for low recoil energies. If the energy ROI is to be widened, these studies have to be extended as well. The energy scale is determined either directly by using mono-energetic neutron sources or by comparing measured neutron energy with simulations. While the former method is more robust, only a fairly small amount of possible neutron sources can be used. Monte Carlo simulations can be used over a wider energy range, but they require additional assumptions and have large uncertainties. Currently energies only up to about 76 keV nuclear recoil energy are calibrated, although with good accuracy~\cite{Akerib:2016mzi}.
Presently the highest energy calibrations performed are using D--D neutron generators, providing neutrons of about 2.5~MeV, thus leading to maximal recoils in liquid xenon of about $76$~keV~\cite{Akerib:2016mzi}.  D--T neutron generators could provide much higher energies, up to about 14.1~MeV, enabling the calibration of nuclear recoils up to an energy of approximately 430~keV in a xenon detector which is very well suited to the extended search window which we propose in this work. Argon, because of its lower atomic mass, can use the same sources to calibrate recoils of about 230 and 1300~keV, respectively~\cite{Polosatkin:2014dka}.

In this analysis, we  consider two simplified xenon and argon experimental setups, shown in table \ref{tab:experiments}, which are motivated by future detectors such as LZ, XENONnT, PandaX, and DARWIN \cite{Akerib:2015cja,Aprile:2017aty,Cui:2017nnn,Aalbers:2016jon} (for Xe), as well as DarkSide and DEAP \cite{Aalbers:2016jon,Amaudruz:2017ekt} (for Ar). For each of these setups, we have adopted two configurations: a {\em nominal} range for the energy ROI, based on current specifications, and an {\em extended} ROI motivated by the possible improvements in calibrating/reducing the high energy background, as explained in the previous section.

\begin{table}[t!]
\centering
\begin{tabular}{|c|c|c|c|}
 \hline
	 Target&Exposure [ton yr]&Nominal ROI [keV]& Extended ROI [keV]\\
	\hline
	 Xe& $15.3$ & $3-30$ & $3-500$\\
    Ar & $20$ & $5-50$ & $5-300$\\
	\hline
	\end{tabular}
\caption{Specifications of the xenon and argon experiments considered in this work inspired by future LZ and DarkSide experiments, respectively.}
\label{tab:experiments}
\end{table}

In this work we adopt the nuclear responses from ref.~\cite{Fitzpatrick:2012ix}, where a comprehensive set of responses is presented. Notice that when considering nuclear recoils at high energies, such approximate nuclear responses can be subject to uncertainties \cite{Vietze:2014vsa}. There are also chiral effective theory effects which can induce corrections to the recoil spectra due to inter-nucleon interactions mediated by meson exchange \cite{Bishara:2016hek,Hoferichter:2015ipa}.  These can, in turn, alter the shape of the nuclear responses \cite{Hoferichter:2016nvd}. We have not included these uncertainties in our analysis. On the one hand, they would not alter the end point of the nuclear recoil spectrum as this only depends on kinematics. On the other hand, the changes in the shape can be relevant for momentum-dependent operators. Determining how much these uncertainties translate into uncertainties on the exclusion limits or parameter reconstruction is beyond the scope of this work. Nonetheless, this work can be seen as a motivation to better understand such uncertainties.

\section{Benefits of enlarging the energy window}
\label{sec:benefits}

In this section we provide concrete examples that illustrate the advantages of enlarging the energy window in the search for DM signals. We will address the effect that a wider ROI has on deriving exclusion limits for EFT operators if no DM signal is found, on the reconstruction of DM parameters in the event of a positive signal, and on gaining sensitivity to the astrophysical parameters describing the DM halo.

\subsection{Exclusion limits}
\label{sec:limits}

The first advantage of increasing the energy range in direct detection data analysis is to obtain better upper bounds on the DM-nucleus cross section, if no excess over the background is found. This argument strongly relies on the expected recoil spectrum from DM interactions and therefore it varies significantly for different EFT operators.

In particular, for the canonical SI and SD ($\op{1}$ and $\op{4}$, respectively) the nuclear recoil spectrum has an approximate exponential behaviour as a function of the recoil energy. A similar behaviour is observed for those EFT operators without an explicit momentum dependence, namely $\op{7,8,12}$, although the different form factors induce some variation. In general, if DM interactions are dominated by any of these operators one would expect that most of the DM signal is concentrated at low energies, and thus the usual strategy of lowering the energy threshold to enlarge the ROI would be optimal. On the other hand, the recoil spectrum for operators with an explicit momentum dependence exhibit a characteristic peak at high energies, and vanish when $E_R\to 0$. The position of this peak is shifted to higher energies as the mass of the DM particle increases. The structure of the nuclear form factors at high energy can also induce further features in the recoil spectrum. As we can see in table~\ref{tab:eft}, this applies to the vast majority of EFT operators $\op{3,5,6,9,10,11,13,14,15}$. On top of these, $\op{6}$ depends quadratically on the transferred momentum. For these operators, it is possible that a significant part of the signal lies at large recoil energies and could be missed if the analysis window is not large enough \cite{Aprile:2017aas,Gelmini:2018ogy}.

\begin{figure}[t!]
\centering
\includegraphics[width=\textwidth]{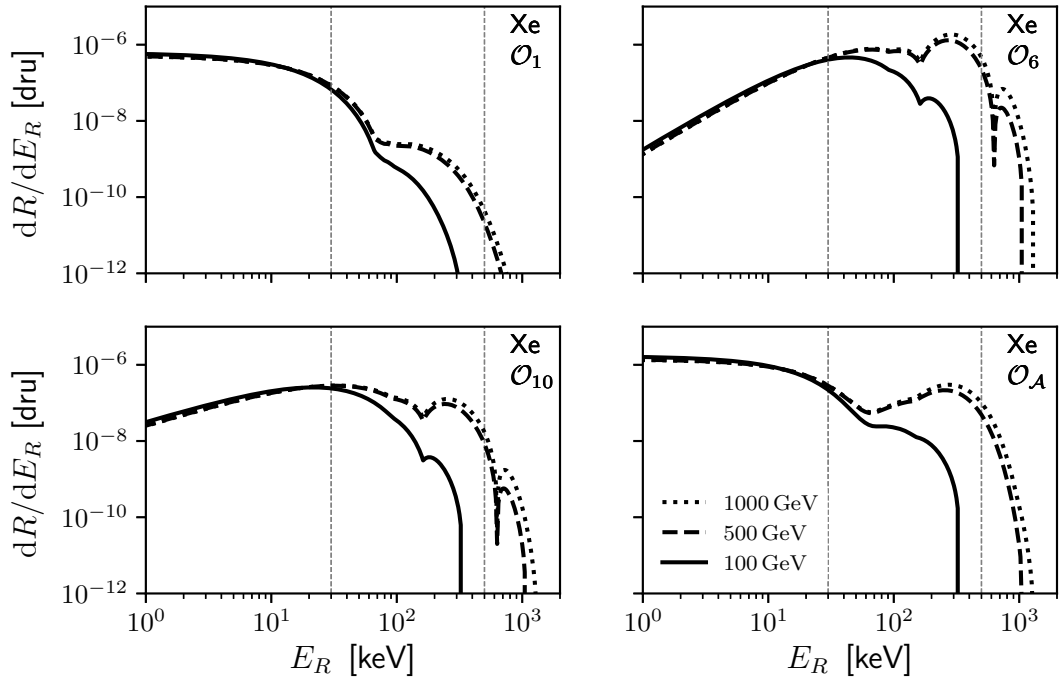}
\caption{The expected recoil spectrum for EFT operators, $\op{1}$ (top left panel), $\op{6}$ (top right panel), $\op{10}$ (bottom left panel), and for anapole interactions (bottom right panel) in a xenon experiment. The DM mass is chosen to be $\mdm=100$~GeV (solid), 500~GeV (dashed), and 1000~GeV (dotted). The vertical dashed lines represent $\Emax=30$~keV and 500~keV. The coupling for each operator has been fixed to produce 100 events in the energy range $[3,\, 30]$~keV.}
\label{fig:spectra}
\end{figure}

To illustrate the discussion above, in figure~\ref{fig:spectra} we compare the recoil spectrum of a typical $q$-independent operator ($\op{1}$) with that of a $q$-dependent operator ($\op{10}$), and a $q^2$-dependent operator ($\op{6}$), as they would be observed in a xenon detector. We also include the non-trivial example of an anapole interaction, which involves a linear combination of operators $\op{8}$ and $\op{9}$. We display the spectra for three DM masses, $m_\chi=100$, $500$, and $1000$~GeV, and fix the couplings such that each example produces 100 nuclear recoil events in the nominal energy range $[3,\,30]$~keV. 
The vertical dashed lines represent the maximum energy in the nominal and extended ROI cases, namely $\Emax=30$ and $500$~keV, respectively.

Irrespectively of the EFT operator, the DM spectrum for elastic scattering displays a maximum energy as a function of the DM escape speed from the Galaxy, $v_{\rm esc}$, and the DM and target nucleus masses, 
\begin{equation}
E_R^{\rm end}=2\,\frac{m_\chi^2m_N}{(m_\chi+m_N)^2}v_{esc}^2\, .
\label{eq:emax}
\end{equation}
For large DM masses, the maximum energy is  a function of the target mass, $E_R^{\rm end}\approx 2 m_N  v_{esc}^2$. 
Using the SHM parameters, we obtain $E_R^{\rm end}\approx 1600$~keV for a xenon target and $E_R^{\rm end}\approx500$~keV for an argon target. 

\begin{figure}[t!]
\centering
\includegraphics[width=\textwidth]{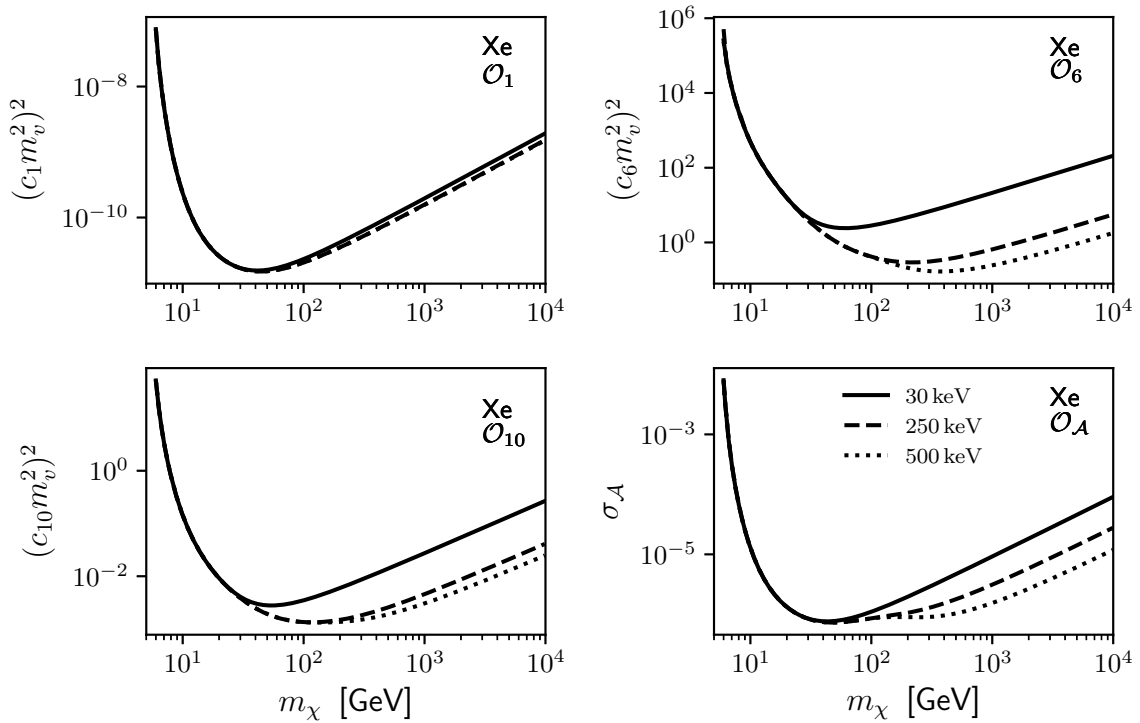}
\caption{Projected exclusion limits for a xenon detector for EFT operators, $\op{1}$, $\op{6}$, $\op{10}$, and for anapole interactions. The solid, dashed, and dotted lines correspond to $\Emax=30,\, 250$ and 500~keV, respectively. }
\label{fig:op6_exclusion_xe}
\end{figure}

As we can observe, if $\Emax$ is increased, a significant part of the DM signal for momentum-dependent operators can be accessed, especially for heavy DM particles, with a more substantial improvement for $\op{6}$. In the case of anapole interactions, an enlarged energy range would allow us to probe the region where the momentum-independent $\op{8}$ dominates over the momentum-dependent $\op{9}$ (which displays a bump at large energies).

In figure~\ref{fig:op6_exclusion_xe} we present the upper limits at $90$\% confidence level on the $c_1^2$, $c_6^2$ , and $c_{10}^2$ coefficients, as well as on the coupling of the anapole moment ${\cal A}$, assuming no DM signal in a xenon detector with an exposure of $15.3$~ton~yr (as given in table~\ref{tab:experiments}). The solid, dashed, and dotted lines show the results for $\Emax=30,\, 250$ and 500~keV, respectively. As we can observe, the improvement for momentum-independent operators (such as $\op{1}$) is negligible, whereas momentum-dependent operators greatly benefit from the increased energy range. In the case of $\op{6}$, the exclusion limit can improve by more than one order of magnitude for DM masses above $\mdm\approx 300$~GeV. We have explicitly checked that in the case of a xenon detector, the improvement in the exclusion limits that one obtains when $\Emax$ increases from 500 keV to 1600~keV is minimal, and therefore the optimal value of the maximum recoil energy is $\Emax \approx 500$~keV.

\begin{figure}[t!]
\centering
\includegraphics[width=0.48\textwidth]{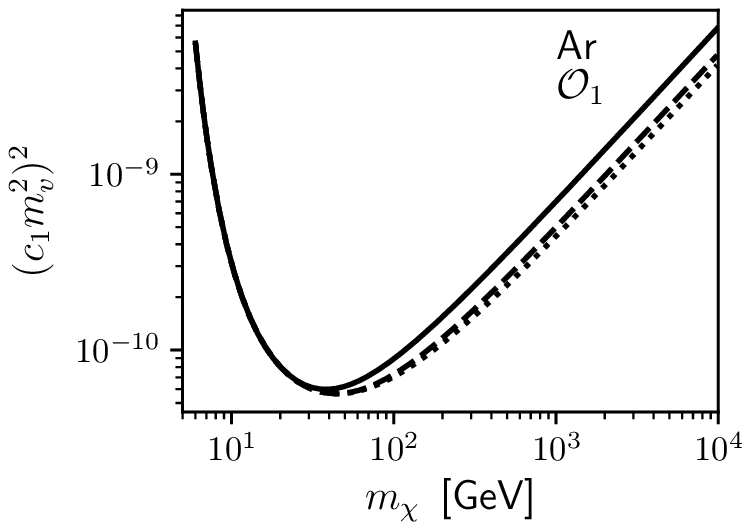}
\includegraphics[width=0.48 \textwidth]{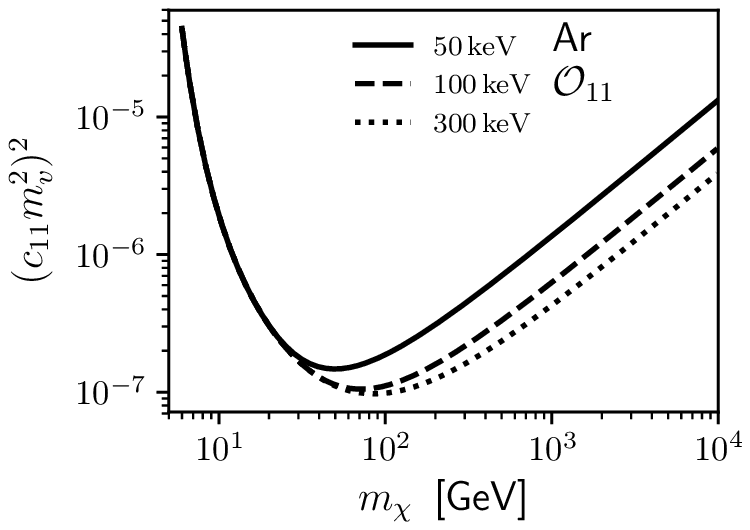}
\caption{Projected exclusion limits for an argon detector for EFT operators, $\op{1}$ and $\op{11}$. The solid, dashed, and dotted lines correspond to $\Emax=50,\, 100$ and 300~keV, respectively. }
\label{fig:op11_exclusion_ar}
\end{figure}

We obtain qualitatively similar results for an argon detector. Figure~\ref{fig:op11_exclusion_ar} shows the exclusion limits obtained for operators $\op{1}$ and the $q$-dependent $\op{11}$. Notice that argon is insensitive to $\op{6}$ and $\op{10}$. We present the results for $\Emax=50$, $100$, and $300$~keV. Once again, the improvement in momentum-independent operators is marginal, but the sensitivity for momentum-dependent ones is greatly enhanced. The improvement for higher values of $\Emax$ is minimal and hence the optimal value of the maximum energy for an argon detector is $\Emax \approx 300$~keV. 

We have checked that the results for other momentum-dependent operators are qualitatively similar to those of $\op{10}$, with small differences that can be attributed to the corresponding form factors.

\subsection{Dark matter parameter reconstruction}
\label{sec:reconstruction}

If an excess compatible with DM is obtained in direct detection experiments, the mass and couplings of the DM particle can be inferred from the nuclear recoil spectrum \cite{Green:2007rb,Green:2008rd,McDermott:2011hx}. From the discussion in the previous section, we can also infer that an increased energy range in the analysis window would lead to a better measurement of the recoil spectrum and, consequently, to a better measurement of the DM parameters.

In particular, the endpoint of the nuclear recoil spectrum, $E_R^{\rm end}$ (see eq.~\ref{eq:emax}), provides a good measurement of the DM mass, irrespectively of the EFT operator. This complements other information that can be obtained from the spectral shape. 
As we can observe in figure~\ref{fig:spectra}, the endpoint for a 100~GeV DM particle is above the canonical energy window considered in xenon experiments, but could be observed with a larger $\Emax$. 
Notice that in the limit of very heavy DM particles, the endpoint is only a function of the target mass and therefore the capability of reconstruction is eventually lost.

We have chosen a set of representative benchmark points and use eq.~(\ref{eq:drate}) to compute the simulated spectrum, $N_k^{\rm obs}$, which we take as the {\it observed} event rate.  We  explore the DM parameter space and for each point, $\lambda=\left\{\mdm,\,c_i\right\}$, in the DM mass and EFT coupling plane, we compute the expected number of DM events in a given energy bin, $N_k(\lambda)$. We then construct the following binned likelihood
\begin{equation}
\mathcal{L}(N^{\textrm{obs}}|{\lambda})=\prod_k \frac{(N^{\rm{th}}_{k})^{N^{\textrm{obs}}_k}}{N^{\textrm{obs}}_k\,!}\exp\left(-N^{\rm{th}}_{k}\right),
\label{eq:loglike}
\end{equation}
where $N_k^{\rm{th}}=N_k(\lambda)+N_k^{\textrm{b}}$ is the total number of counts predicted in bin $k$, taking into account the number of background events $N_k^b$. We assume a bin width of 1~keV in the nominal ROIs. For the extended ROIs we increase the bin size to 50~keV in xenon and 10~keV in argon\footnote{Lacking information about energy calibration in the extended ROI, we have decided to take a conservative approach and increase the bin size. A smaller bin size could in principle lead to a better parameter reconstruction.}. In order to calculate the profile likelihood and to effectively scan the parameter space, we use MultiNest \cite{Feroz:2008xx}. The confidence intervals in the parameter space are then extracted using Superplot \cite{Fowlie:2016hew}. To speed up the computation of the number of DM events in the EFT framework, we use the surrogate model RAPIDD \cite{Cerdeno:2018bty}.

In order to quantify the improvement in the DM mass reconstruction when the energy range is extended, we simulate a future DM excess, assuming a given EFT operator and DM mass.  Then we attempt to reconstruct it using the binned likelihood defined in eq.~(\ref{eq:loglike}). By construction, the best fit point coincides with the simulated DM mass, but we also determine the $1\sigma$ confidence interval of the reconstructed masses, $(m_{1\sigma}^-,\,m_{1\sigma}^+)$, from which we define 
\begin{equation}
	\Delta_m = \left( \frac{m_{1\sigma}^-}{\mdm},\,\frac{m_{1\sigma}^+}{\mdm}\right) \ .
\label{eq:deltam}
\end{equation}
By construction, the true value is at $\Delta_m=1$ and $\Delta_m\in[0,\,\infty]$.
For concreteness, we will consider benchmark points that predict observation of 100 DM-induced nuclear recoil events.

\begin{figure}[!t]
\centering
\hspace*{-0.25cm}
\includegraphics[width=0.35\textwidth]{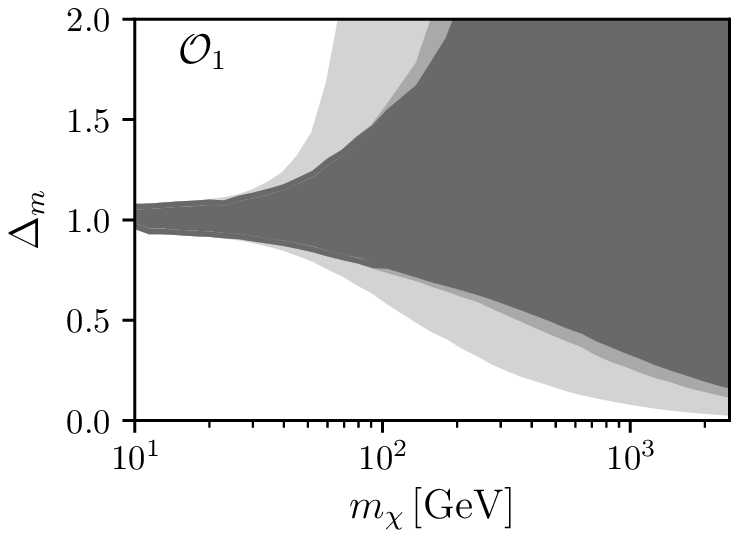}\hspace*{-0.25cm}
\includegraphics[width=0.35\textwidth]{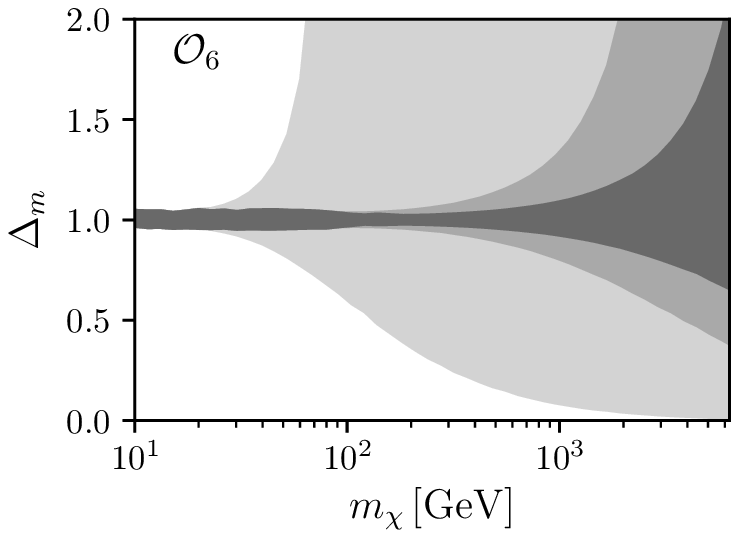}\hspace*{-0.25cm}
\includegraphics[width=0.35\textwidth]{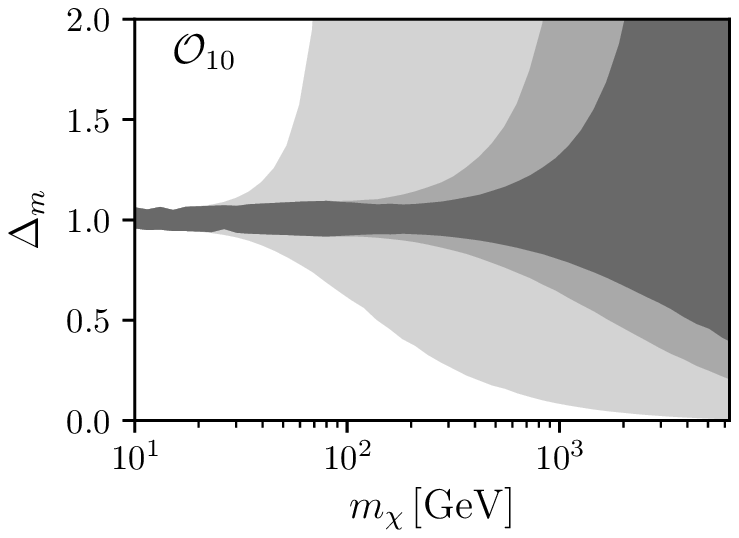}
\caption{Mass reconstruction parameter, $\Delta_m$ (defined in eq.~\ref{eq:deltam}), as a function of the DM mass for operators $\op{1}$, $\op{6}$, and $\op{10}$ from the left to right panels in a xenon detector. We have assumed a benchmark point, for each value of the DM mass, that produces 100 nuclear recoil events. The  light to dark grey regions correspond to different energy ROIs, with $\Emax=30,\,250\,, 500$~keV, respectively.} 
\label{fig:deltam_100}
\end{figure}

Figure~\ref{fig:deltam_100} shows the resulting $\Delta_m$ as a function of the DM mass for a reconstruction using 100 events for each value of the DM mass. From the left to right panels, we present the results for EFT operators $\op{1}$, $\op{6}$, and $\op{11}$. The different coloured areas correspond to different energy ROIs, with $\Emax=30$, $250$, and $500$~keV.

Let us first concentrate on the canonical spin-independent operator $\op{1}$. For the nominal ROI, with $\Emax=30$~keV the $1\sigma$ region is unbounded from above for DM masses above 60~GeV. As $\Emax$ increases, larger DM masses can be accessed and with $\Emax=250$~keV one can successfully reconstruct DM masses up to 100~GeV. The optimal energy range of $\Emax=500$~keV would allow us to reconstruct DM masses up to just 200~GeV. For any DM mass above this value, there is no upper bound in the reconstructed value. Interestingly, the lower limit of the reconstruction also benefits from a larger $\Emax$. It should also be noted that the relative improvement in reconstruction from having a maximum energy above  500~keV is minimal. The reconstruction for 100 nuclear recoil events shows that the limitation in the reconstruction of DM masses is not due to poor statistics. Indeed, even with a larger number of events, there is no proper DM mass reconstruction for $\op{1}$ above $\mdm\sim200$~GeV.

The benefit of enlarging the energy ROI is  more pronounced for momentum-dependent operators, such as $\op{6}$ and $\op{10}$. In both cases, increasing $\Emax$ leads to a much better DM mass reconstruction, where DM masses as heavy as $\mdm\sim2$~TeV can be resolved. This is in contrast with the nominal ROI, which only allows reconstruction up to $\mdm\sim60$~GeV.
It should be noted that even when the upper limit of the reconstructed mass becomes unbounded ($\Delta_m\to\infty$) the lower limit can still improve substantially if $\Emax$ is further increased.

\begin{figure}[t!]
\centering
\includegraphics[width=0.85\textwidth]{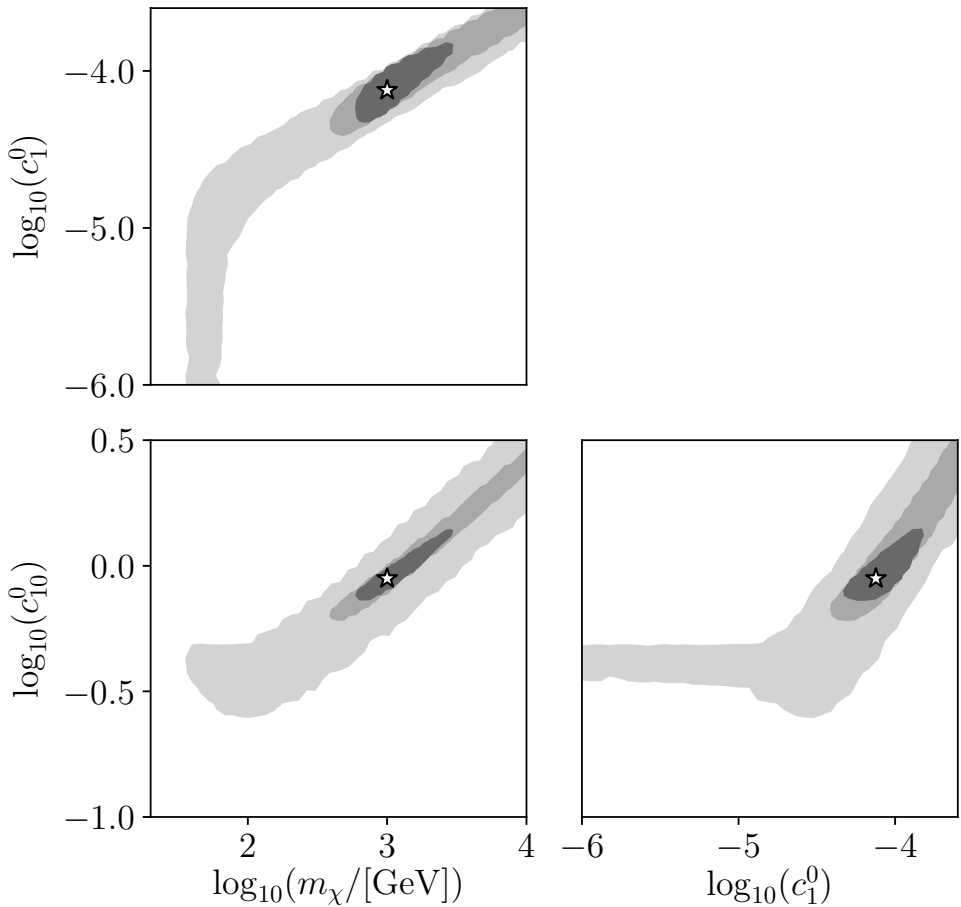}
\caption{Results from a 3D parameter scan, fitting mock data from our benchmark point with mass $m_\chi=1$ TeV and a coupling to $\op{1}$ and $\op{10}$ which produces 100 counts in the $[3, 30]$~keV window. The $2\sigma$ countours are shown from light gray to dark gray for $\Emax=30,\, 250$ and $500$ keV, respectively. The white star represents the benchmark point.}
\label{fig:3Dscan}
\end{figure}

Having a good reconstruction of the DM mass also helps in measuring the DM couplings and removing degeneracies in the parameter space. We illustrate this with an example, motivated by the case of scalar DM with a scalar mediator, in which the parameter space is three-dimensional and consists of the DM mass and operators $\op{1}$ and $\op{10}$. We have selected a benchmark point with $\mdm=1$~TeV and equal contributions from $\op{1}$ and $\op{10}$ that give 100 counts in the usual $[3,30]$~keV energy window. This benchmark point is  special because, as we saw in the discussion above, $\mdm$ values can be bounded from above in the extended energy ROI. Figure~\ref{fig:3Dscan} represents the $2\sigma$ reconstructed region in the 3D parameter space $(\mdm,\,c^0_1,\,c_{10}^0)$ for the three choices $\Emax=30$, $250$, and $500$~keV. As we can observe, the nominal ROI is insufficient to determine any of the three parameters and large degeneracies are observed in the three planes. 
For example, the data can be consistent with light DM that interacts mainly through $\op{10}$ or with heavier DM and a linear combination of $\op{1}$ and $\op{10}$. When the energy ROI is extended, the light DM solution disappears and eventually, with $\Emax=500$~keV, a full reconstruction of the three parameters is possible.

\begin{figure}[t!]
\centering
\includegraphics[width=0.85\textwidth]{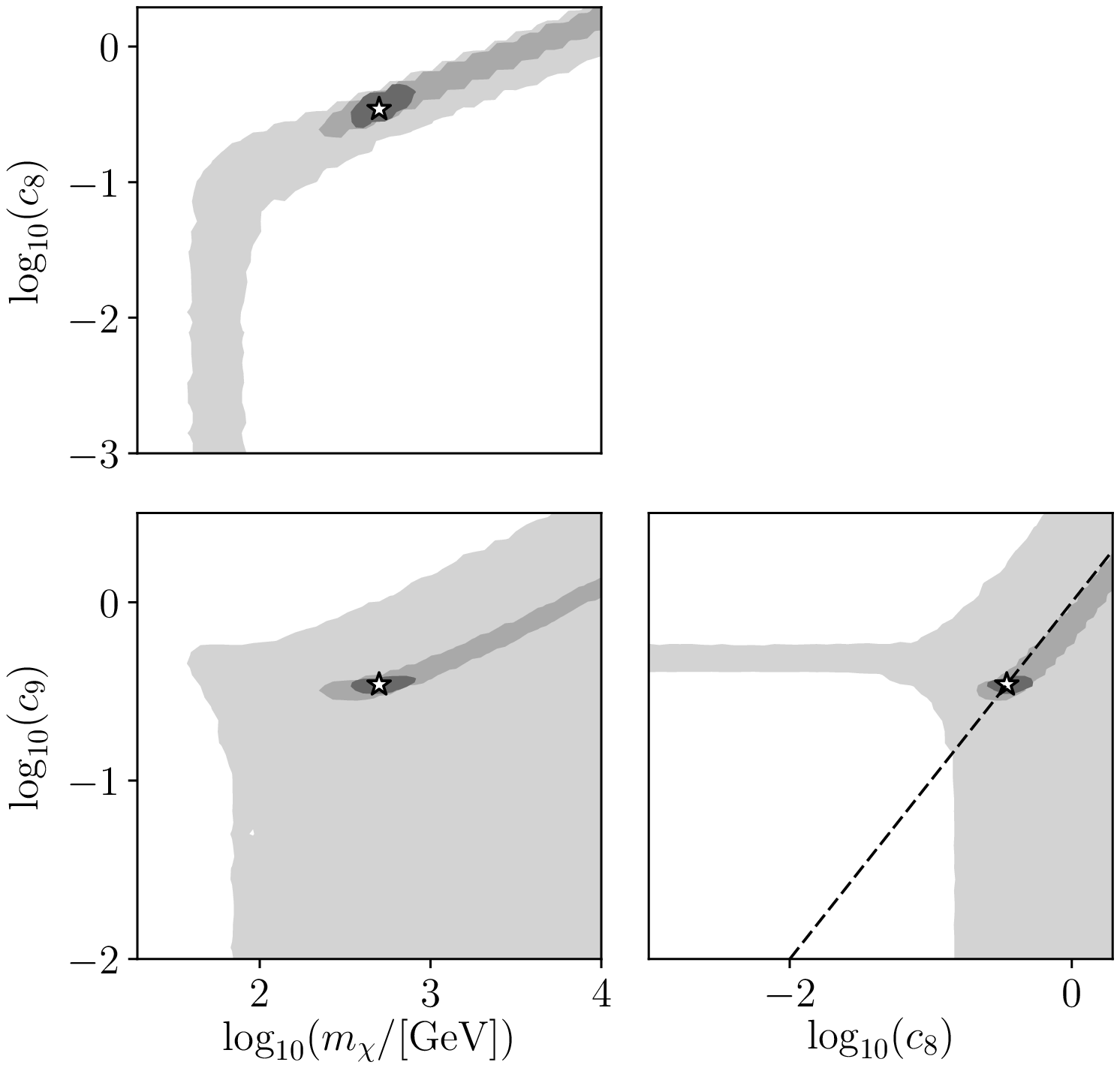}
\caption{Same as figure \ref{fig:3Dscan} but for $m_\chi=500$~GeV and a coupling via the anapole moment. The black dashed line in the $(c_9,c_8)$ plane represents the relation among these couplings in an anapole DM model.} 
\label{fig:3Dscan_anapole}
\end{figure}

As a second example, we have selected a benchmark point that corresponds to anapole DM, with $\mdm=500$~GeV and $c_8=c_8^p=2\mathcal{A}e$, $c_8^n=0$, $c_9=c_8=-(2/g_p)c_9^p=-(2/g_n)c_9^n$. For simplicity we have already imposed the correct relation among the couplings to protons and neutrons and we have attempted to reconstruct the DM parameters in the 3D plane $(\mdm,\,c_8,\,c_{9})$. The results are shown in figure~\ref{fig:3Dscan_anapole}, and in the $(c_8,\,c_9)$ plane we indicate the anapole relation between both couplings by means of a dashed black line. As in the previous example, the results with the nominal ROI are not sufficient to provide a good measurement of any of the parameters and large degeneracies are visible. For $\Emax=250$~keV, the region with low DM mass and a leading $\op{9}$ contribution disappears and although there is still a residual degeneracy in the $(c_8,\,c_9)$ plane, the results already reproduce the anapole relation along the dashed line. Notice that, from the spectrum in the bottom right panel of figure~\ref{fig:spectra}, we can see that this energy is enough to observe the characteristic shape of $\op{9}$ but it is insufficient to measure the end point of the spectrum. With the extended ROI ($\Emax=500$~keV), a good measurement is obtained for the three parameters. Without enlarging the energy ROI, it would be impossible to properly identify this scenario as anapole DM.

\subsection{Astrophysical uncertainties}
\label{sec:astro}

One of the major sources of uncertainty which enters in the predicted event rate is the uncertainty in the local DM distribution. In particular, changes in the DM density and velocity distribution in the Solar neighborhood lead to large variations in the allowed regions or exclusion limits obtained in the plane of DM-nucleon coupling and DM mass from direct detection experiments.

An important way to obtain information on the DM velocity distribution is to extract it from cosmological simulations. High resolution cosmological simulations including baryonic physics have recently become available and are able to reproduce important galactic properties with significant agreement with observations. Recently it was shown that the local DM velocity distribution extracted from state-of-the-art hydrodynamic simulations fits well a Maxwellian distribution, but with a peak speed which can be different from the local circular speed~\cite{Bozorgnia:2016ogo, Kelso:2016qqj, Sloane:2016kyi,Bozorgnia:2017brl}. Ref.~\cite{Bozorgnia:2016ogo} used the EAGLE and APOSTLE high resolution simulations which include both DM and baryons and identified 14 simulated {\it Milky Way-like} galaxies by taking into account observational constraints on the Milky Way. The range of the best fit peak speeds of
the Maxwellian distribution for the simulated Milky Way-like galaxies was found to be $223-289$ km$/$s. To include in the analysis of direct detection data the largest possible deviation with respect to the SHM predicted by simulations, we will consider the simulated MW-like galaxy in EAGLE/APOSTLE with the local DM velocity distribution furthest from the SHM. The parameters of this halo, along with the fiducial parameters of the SHM are given in table~\ref{tab:astroparam}. For both halo models, we consider the velocity of the Earth with respect to the Sun and the peculiar velocity of the Sun as given in ref.~\cite{Bozorgnia:2016ogo}.

\begin{table}[t!]
    \centering
    \begin{tabular}{|c|c|c|c|c|}
      \hline
        Parameter & $v_{\textrm{peak}}$ [km s$^{-1}$]  & $v_c$ [km s$^{-1}$]  & $v_{\textrm{esc}}$ [km s$^{-1}$] & $\rho_0$ [GeV cm$^{-3}$]  \\
       \hline
       SHM &220  & 220& 544 & 0.4 \\
       EAGLE &288.64 & 254.06& 874.76  &0.68\\
      \hline
    \end{tabular}
    \caption{The peak speed of the Maxwellian velocity distribution, local circular speed, Galactic escape speed, and the local DM density assumed in the SHM (row 1) and extracted from the simulated MW-like galaxy in the EAGLE simulation farthest from the SHM (row 2).}
    \label{tab:astroparam}
  \end{table}

For the current recoil energy range probed by direct detection experiments, uncertainties in the high velocity tail of the DM velocity distribution become especially important for light DM masses. This is because for a fixed maximum recoil energy, low DM masses lead to a high minimum DM speed, $v_{\rm min}$, where the experiments probe the tail of the DM velocity distribution. In particular, the range of the best fit peak speeds for the Maxwellian velocity distribution allowed by hydrodynamic simulations of MW-like galaxies~\cite{Bozorgnia:2016ogo,Bozorgnia:2017brl} translates into an uncertainty in the allowed regions or exclusion limits set by direct detection experiments, for low mass (10's of GeV) DM particles. However, if the maximum recoil energy in an experiment is significantly increased, even larger DM masses (100's of GeV) lead to high $v_{\rm min}$, where the experiments become sensitive to the tails of the DM velocity distribution.

In order to disentangle astrophysical uncertainties, one has to consider results from multiple targets. As discussed extensively in the literature, this complementarity will also help with particle parameter reconstruction \cite{Bertone:2007xj,Pato:2010zk,Cerdeno:2013gqa,Peter:2013aha,Catena:2014uqa, Edwards:2018lsl}. Hence, we will once again consider future Xe and Ar detectors. To determine whether increasing the energy window will help us overcome astrophysical uncertainties for large DM masses, we produce a series of data resulting in 100 and 1000 counts in the two detector configurations given in table \ref{tab:experiments}, for a 100 GeV DM, and arising from the two halo models (SHM and EAGLE) we are considering, which we take as the {\it true} halo model. We then perform a parameter reconstruction for each experiment separately, assuming the SHM for both cases. If one was to observe some tension between the two experiments, it would suggest that the assumption of the SHM would be incorrect.

\begin{figure}[t!]
\centering
\includegraphics[width=\textwidth]{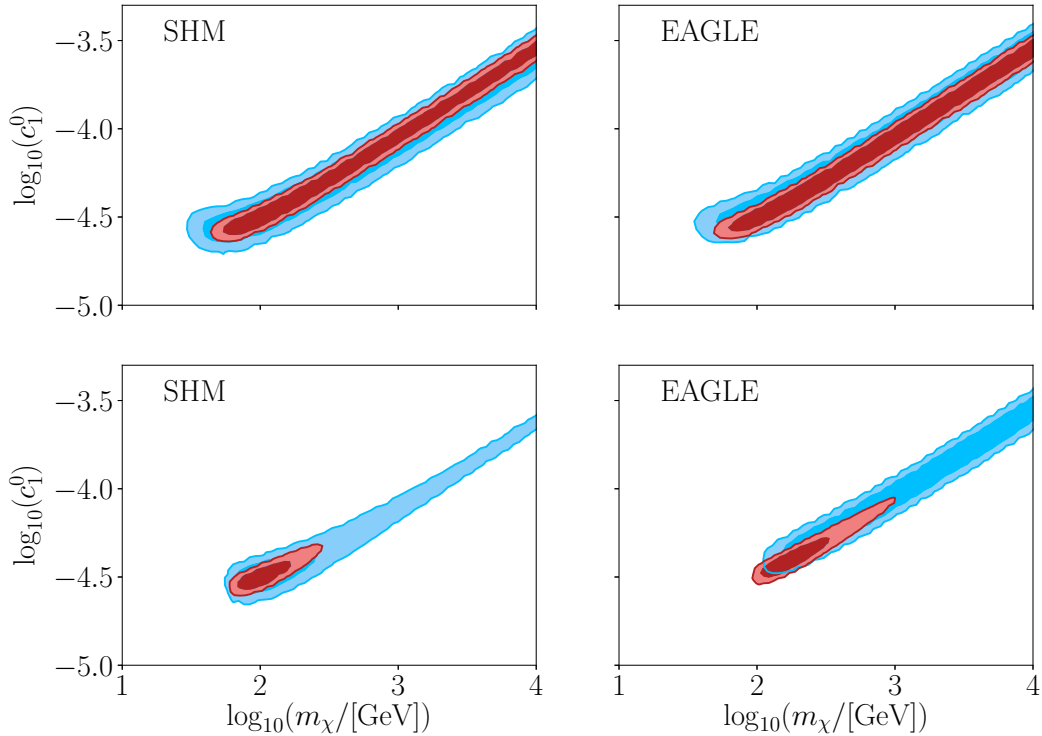}
\caption{Profile likelihoods for the reconstruction of DM parameters for a 100 GeV DM benchmark point that produces 100 counts in our future Xe and Ar detector configurations given in table~\ref{tab:experiments}, for a coupling only to $\op{1}$. Red and blue shaded regions correspond to $2\sigma$ (light shade) and $1\sigma$ (dark shade) regions obtained from the individual targets of Xe and Ar, respectively. The top row corresponds to the {\it nominal} configuration of the experiments assuming an energy window of $[3,\,30]$~keV for Xe and $[5,\,50]$~keV for Ar. The bottom row corresponds to the {\it extended} configuration assuming an energy window of $[3,\,500]$~keV for Xe and $[5,\,300]$~keV for Ar. The left and right panels correspond to the SHM and EAGLE halo models, respectively, which were assumed to generate the data (i.e.~the {\it true} halo model). Every parameter scan has been performed assuming the SHM with values given in table~\ref{tab:astroparam}.}
\label{fig:Op1_100GeV_100cs}
\end{figure}

Figure~\ref{fig:Op1_100GeV_100cs} shows the profile likelihood for the reconstruction of DM parameters for a 100~GeV benchmark point that produces 100~counts in our Xe (red) and Ar (blue) detector configurations using the {\it nominal} (top row) and {\it extended} (bottom row) ROIs given in table~\ref{tab:experiments}, for a coupling to $\op{1}$. It is clear from figure \ref{fig:Op1_100GeV_100cs} that by opening the energy window for both experiments, we achieve stronger constraints on the coupling and DM mass and this in turn can put tension between the two experiments for the EAGLE halo model. However, we see that with 100 counts this tension is not very strong, and it would be difficult to rule out the SHM at high significance.

There are some general comments to be made. If the halo assumed in the reconstruction of the DM parameters has a $v_{\textrm{esc}}$ below the {\it true} value, then the spectra will appear to come from a DM particle with a larger $\mdm$. Notice that the smaller the mass of the target nucleus, the greater the effect. Furthermore, for $\mdm \gg m_N$, the higher the value of $\mdm$, the higher is the coupling strength, causing the reconstructed region to move upwards. This can be seen in figure \ref{fig:Op1_100GeV_100cs}, where the reconstructed region for the EAGLE halo which has a larger escape speed compared to the SHM, is shifted to masses larger than 100 GeV. This shift is even larger for the Ar experiment which has a smaller target nucleus mass. 
These results complement the findings of ref.~\cite{Peter:2011eu}, where it was found that the reconstructed areas could significantly improve with an extended analysis window. Our results strengthen the complementary role between xenon and argon targets for the study of large mass DM. 

\begin{figure}[t!]
\centering
\includegraphics[width=\textwidth]{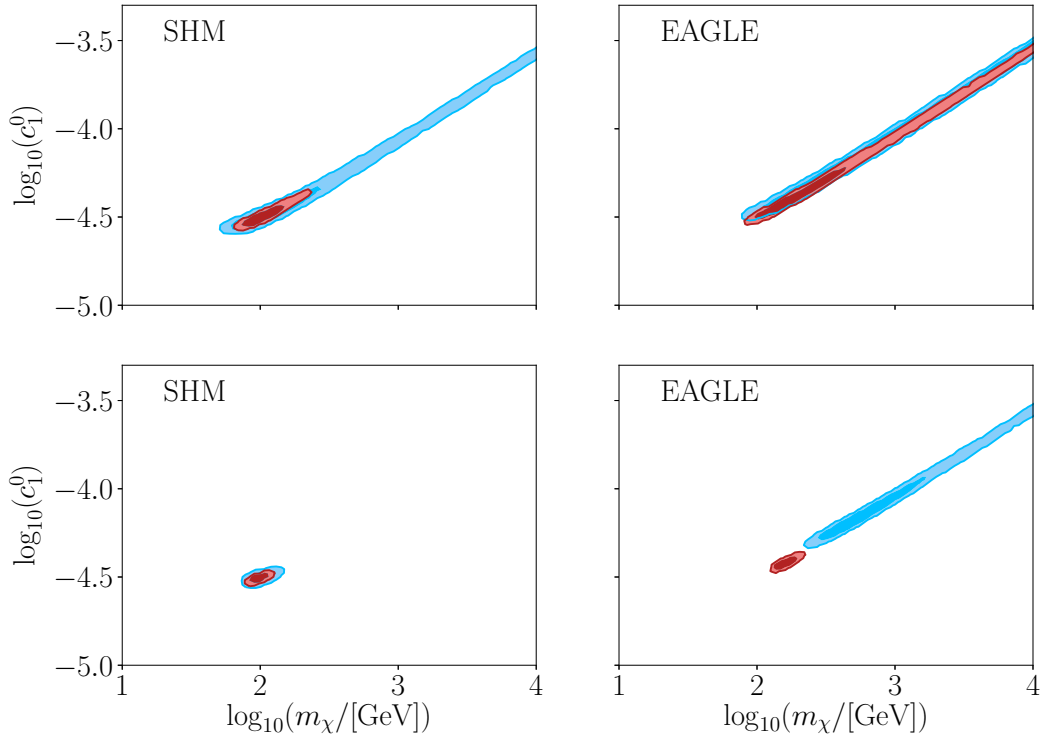}
\caption{Same as figure \ref{fig:Op1_100GeV_100cs}, but with an experimental exposure increased by a factor of 10, i.e.~a benchmark point that produces 1000 counts.}
\label{fig:Op1_100GeV_1000cs}
\end{figure}

In order to see whether greater statistics would improve our ability to distinguish between different halos we ran the analysis again but with ten times the original exposure given in table \ref{tab:experiments} for both the Xe and Ar experiments. To do this, we had to include the neutrino floor as a known background into the calculation. Figure \ref{fig:Op1_100GeV_1000cs} shows the results for the $\op{1}$ benchmark that produces a 1000 counts. We see that for both halos, the different experiments are completely consistent in the  nominal setup. Once we open the energy window we start to have tension between different experiments for the EAGLE halo. In particular, the $2\sigma$ regions for the different detectors are completely separated for the EAGLE halo, and hence the SHM could be ruled out in this case.

\begin{figure}[t!]
\centering
\includegraphics[width=\textwidth]{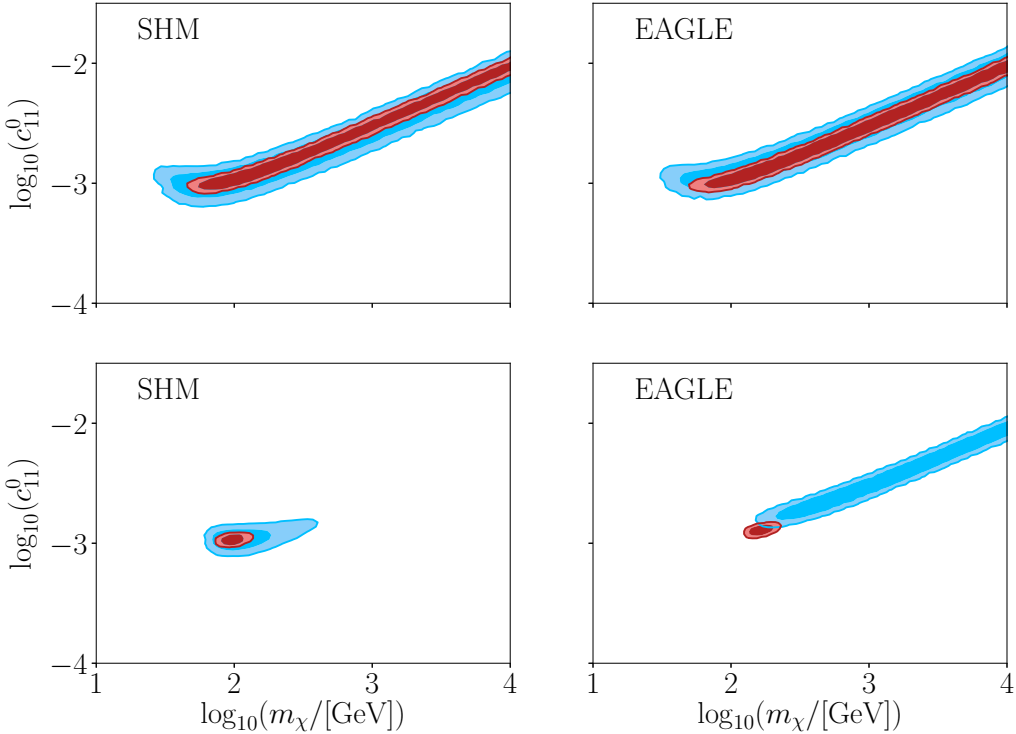}
\caption{Same as figure \ref{fig:Op1_100GeV_100cs} but with a coupling only to operator $\op{11}$.}
\label{fig:Op11_100GeV_100cs}
\end{figure}

As an example of an operator with a different behavior with respect to $\op{1}$, we next consider the $\op{11}$ operator which exhibits a $q$-dependence. Figure~\ref{fig:Op11_100GeV_100cs} shows the results for coupling to $\op{11}$ for a 100 GeV benchmark point that produces 100 counts in Xe and Ar. We can see that even with a 100 counts, when the energy window is opened, $\op{11}$ interactions can cause greater tension between the two experiments for the EAGLE halo, resulting in separated $1\sigma$ regions for Xe and Ar.

\section{Conclusions}
\label{sec:Conclusion}

In this article, we have investigated the benefits of enlarging the region of interest of the nuclear recoil energy in the analysis of direct DM detection data. In particular, we have studied how increasing the maximum recoil energy, $\Emax$, will allow us to extract more information from the tail of the nuclear recoil spectrum. We have concentrated on elastic DM-nucleus scattering and considered an EFT approach to describe the interaction. Focusing on future xenon and argon detectors, we have investigated the implications that a larger $\Emax$ has on setting limits on DM couplings, on measuring the DM parameters, and on obtaining information about the astrophysical parameters of the DM halo.

In agreement with previous studies, we find that a larger $\Emax$ would lead to more stringent upper bounds on the DM-nucleus cross section, especially for momentum-dependent operators. For example, in xenon experiments, the sensitivity to $\op{6}$ and $\op{10}$ can increase by approximately an order of magnitude for DM masses above approximately 300~GeV, and in argon a similar improvement can be achieved for $\op{11}$. We have determined that the optimal values of the maximum energy are $\Emax\approx500$~keV for xenon and $\Emax\approx300$~keV for argon. This would require a good knowledge of the experimental background at those energies, as well as proper calibration (that could be achieved using a D--T source).

We have also studied how well the mass and couplings of the DM particle could be reconstructed from future data if an excess over the background is observed. We point out that a larger $\Emax$ would allow for a much better measurement of the DM mass, mainly from the observation of the end-point of the nuclear recoil spectrum. For momentum-independent operators, the improvement is modest. For example, in the case of $\op{1}$ the extended ROI allows to reconstruct DM masses up to approximately 200~GeV. However, the improvement for momentum-dependent operators is much more impressive. For example, for operator $\op{6}$ DM masses as large as 2~TeV can be reconstructed with an extended ROI.

We have also investigated three-dimensional parameter reconstruction, concentrating on two examples: scalar DM with scalar mediator, and anapole DM. Our results show that opening the energy ROI is an excellent way to identify the linear combination of momentum-dependent and momentum-independent operators. This is crucial to distinguish different DM models in the range of heavy DM masses.

Finally, we have shown that an extended energy ROI can also be used to test astrophysical parameters of the DM halo, such as the DM escape speed. In a series of examples, we have simulated direct detection data using two different halo models and attempted to reconstruct the DM couplings and mass using the Standard Halo Model. We have observed that an extended ROI and target complementarity can help in identifying when the wrong hypothesis is used in the data analysis.

\vspace*{1cm}
\noindent{\Large {\bf Acknowledgements}}

\noindent We are grateful to Bradley Kavanagh, Alexis Plascencia, Ibles Olcina and Elliott Reid  for useful discussions. NB acknowledges support from the COFUND Junior Research Fellowship. BP is grateful for the support of Brandeis University. AC is grateful for the support from the Science and Technology Facilities Council (STFC). 

\bibliographystyle{JHEP-cerdeno}
\bibliography{library}

\providecommand{\href}[2]{#2}\begingroup\raggedright\begin{thebibliography}{10}

\bibitem{Goodman:1984dc}
M.~W. Goodman and E.~Witten, \emph{{Detectability of Certain Dark Matter
  Candidates}}, \href{http://dx.doi.org/10.1103/PhysRevD.31.3059}{\emph{Phys.
  Rev.} {\bf D31} (1985) 3059}.

\bibitem{Primack:1988zm}
J.~R. Primack, D.~Seckel and B.~Sadoulet, \emph{{Detection of Cosmic Dark
  Matter}},
  \href{http://dx.doi.org/10.1146/annurev.ns.38.120188.003535}{\emph{Ann. Rev.
  Nucl. Part. Sci.} {\bf 38} (1988) 751--807}.

\bibitem{Aprile:2018dbl}
{\scshape XENON} collaboration, E.~Aprile et~al., \emph{{Dark Matter Search
  Results from a One Tonne$\times$Year Exposure of XENON1T}},
  \href{http://arxiv.org/abs/1805.12562}{{\tt 1805.12562}}.

\bibitem{Akerib:2017kat}
{\scshape LUX} collaboration, D.~S. Akerib et~al., \emph{{Limits on
  spin-dependent WIMP-nucleon cross section obtained from the complete LUX
  exposure}},
  \href{http://dx.doi.org/10.1103/PhysRevLett.118.251302}{\emph{Phys. Rev.
  Lett.} {\bf 118} (2017) 251302}, [\href{http://arxiv.org/abs/1705.03380}{{\tt
  1705.03380}}].

\bibitem{Cui:2017nnn}
{\scshape PandaX-II} collaboration, X.~Cui et~al., \emph{{Dark Matter Results
  From 54-Ton-Day Exposure of PandaX-II Experiment}},
  \href{http://dx.doi.org/10.1103/PhysRevLett.119.181302}{\emph{Phys. Rev.
  Lett.} {\bf 119} (2017) 181302}, [\href{http://arxiv.org/abs/1708.06917}{{\tt
  1708.06917}}].

\bibitem{Agnes:2018fwg}
{\scshape DarkSide} collaboration, P.~Agnes et~al., \emph{{DarkSide-50 532-day
  Dark Matter Search with Low-Radioactivity Argon}},
  \href{http://arxiv.org/abs/1802.07198}{{\tt 1802.07198}}.

\bibitem{Agnese:2018gze}
{\scshape SuperCDMS} collaboration, R.~Agnese et~al., \emph{{Search for
  Low-Mass Dark Matter with CDMSlite Using a Profile Likelihood Fit}},
  \href{http://arxiv.org/abs/1808.09098}{{\tt 1808.09098}}.

\bibitem{Petricca:2017zdp}
{\scshape CRESST} collaboration, F.~Petricca et~al., \emph{{First results on
  low-mass dark matter from the CRESST-III experiment}},  in \emph{{15th
  International Conference on Topics in Astroparticle and Underground Physics
  (TAUP 2017) Sudbury, Ontario, Canada, July 24-28, 2017}}, 2017.
\newblock \href{http://arxiv.org/abs/1711.07692}{{\tt 1711.07692}}.

\bibitem{Aguilar-Arevalo:2016ndq}
{\scshape DAMIC} collaboration, A.~Aguilar-Arevalo et~al., \emph{{Search for
  low-mass WIMPs in a 0.6 kg day exposure of the DAMIC experiment at SNOLAB}},
  \href{http://dx.doi.org/10.1103/PhysRevD.94.082006}{\emph{Phys. Rev.} {\bf
  D94} (2016) 082006}, [\href{http://arxiv.org/abs/1607.07410}{{\tt
  1607.07410}}].

\bibitem{Akerib:2016vxi}
{\scshape LUX} collaboration, D.~S. Akerib et~al., \emph{{Results from a search
  for dark matter in the complete LUX exposure}},
  \href{http://dx.doi.org/10.1103/PhysRevLett.118.021303}{\emph{Phys. Rev.
  Lett.} {\bf 118} (2017) 021303}, [\href{http://arxiv.org/abs/1608.07648}{{\tt
  1608.07648}}].

\bibitem{Dobson:2018xxl}
{\scshape LZ} collaboration, D.~S. Akerib et~al., \emph{{ Projected WIMP
  sensitivity of the LUX-ZEPLIN (LZ) dark matter experiment}}, {\emph{Submitted
  to: Phys. Rev. D.} (2018) }, [\href{http://arxiv.org/abs/1802.06039}{{\tt
  1802.06039}}].

\bibitem{aprile:2015xxl}
{\scshape XENON1T} collaboration, E.~Aprile et~al., \emph{{ Physics reach of
  the XENON1T dark matter experiment}},
  \href{http://dx.doi.org/10.1088/1475-7516/2016/04/027}{\emph{JCAP} {\bf 2016}
  (2016) 36}, [\href{http://arxiv.org/abs/1512.07501}{{\tt 1512.07501}}].

\bibitem{Aalbers:2016jon}
{\scshape DARWIN} collaboration, J.~Aalbers et~al., \emph{{DARWIN: towards the
  ultimate dark matter detector}},
  \href{http://dx.doi.org/10.1088/1475-7516/2016/11/017}{\emph{JCAP} {\bf 1611}
  (2016) 017}, [\href{http://arxiv.org/abs/1606.07001}{{\tt 1606.07001}}].

\bibitem{Aalseth:2017fik}
C.~E. Aalseth et~al., \emph{{DarkSide-20k: A 20 tonne two-phase LAr TPC for
  direct dark matter detection at LNGS}},
  \href{http://dx.doi.org/10.1140/epjp/i2018-11973-4}{\emph{Eur. Phys. J. Plus}
  {\bf 133} (2018) 131}, [\href{http://arxiv.org/abs/1707.08145}{{\tt
  1707.08145}}].

\bibitem{Amaudruz:2017ekt}
{\scshape DEAP-3600} collaboration, P.~A. Amaudruz et~al., \emph{{First results
  from the DEAP-3600 dark matter search with argon at SNOLAB}},
  \href{http://arxiv.org/abs/1707.08042}{{\tt 1707.08042}}.

\bibitem{Fan:2010gt}
J.~Fan, M.~Reece and L.-T. Wang, \emph{{Non-relativistic effective theory of
  dark matter direct detection}},
  \href{http://dx.doi.org/10.1088/1475-7516/2010/11/042}{\emph{JCAP} {\bf 1011}
  (2010) 042}, [\href{http://arxiv.org/abs/1008.1591}{{\tt 1008.1591}}].

\bibitem{Fitzpatrick:2012ix}
A.~L. Fitzpatrick, W.~Haxton, E.~Katz, N.~Lubbers and Y.~Xu, \emph{{The
  Effective Field Theory of Dark Matter Direct Detection}},
  \href{http://dx.doi.org/10.1088/1475-7516/2013/02/004}{\emph{JCAP} {\bf 1302}
  (2013) 004}, [\href{http://arxiv.org/abs/1203.3542}{{\tt 1203.3542}}].

\bibitem{Anand:2013yka}
N.~Anand, A.~L. Fitzpatrick and W.~C. Haxton, \emph{{Weakly interacting massive
  particle-nucleus elastic scattering response}},
  \href{http://dx.doi.org/10.1103/PhysRevC.89.065501}{\emph{Phys. Rev.} {\bf
  C89} (2014) 065501}, [\href{http://arxiv.org/abs/1308.6288}{{\tt
  1308.6288}}].

\bibitem{Dent:2015zpa}
J.~B. Dent, L.~M. Krauss, J.~L. Newstead and S.~Sabharwal, \emph{{General
  analysis of direct dark matter detection: From microphysics to observational
  signatures}}, \href{http://dx.doi.org/10.1103/PhysRevD.92.063515}{\emph{Phys.
  Rev.} {\bf D92} (2015) 063515}, [\href{http://arxiv.org/abs/1505.03117}{{\tt
  1505.03117}}].

\bibitem{Catena:2014epa}
R.~Catena, \emph{{Prospects for direct detection of dark matter in an effective
  theory approach}},
  \href{http://dx.doi.org/10.1088/1475-7516/2014/07/055}{\emph{JCAP} {\bf 1407}
  (2014) 055}, [\href{http://arxiv.org/abs/1406.0524}{{\tt 1406.0524}}].

\bibitem{Schneck:2015eqa}
{\scshape SuperCDMS} collaboration, K.~Schneck et~al., \emph{{Dark matter
  effective field theory scattering in direct detection experiments}},
  \href{http://dx.doi.org/10.1103/PhysRevD.91.092004}{\emph{Phys. Rev.} {\bf
  D91} (2015) 092004}, [\href{http://arxiv.org/abs/1503.03379}{{\tt
  1503.03379}}].

\bibitem{TuckerSmith:2001hy}
D.~Tucker-Smith and N.~Weiner, \emph{{Inelastic dark matter}},
  \href{http://dx.doi.org/10.1103/PhysRevD.64.043502}{\emph{Phys. Rev.} {\bf
  D64} (2001) 043502}, [\href{http://arxiv.org/abs/hep-ph/0101138}{{\tt
  hep-ph/0101138}}].

\bibitem{Graham:2010ca}
P.~W. Graham, R.~Harnik, S.~Rajendran and P.~Saraswat, \emph{{Exothermic Dark
  Matter}}, \href{http://dx.doi.org/10.1103/PhysRevD.82.063512}{\emph{Phys.
  Rev.} {\bf D82} (2010) 063512}, [\href{http://arxiv.org/abs/1004.0937}{{\tt
  1004.0937}}].

\bibitem{Dienes:2014via}
K.~R. Dienes, J.~Kumar, B.~Thomas and D.~Yaylali, \emph{{Dark-Matter Decay as a
  Complementary Probe of Multicomponent Dark Sectors}},
  \href{http://dx.doi.org/10.1103/PhysRevLett.114.051301}{\emph{Phys. Rev.
  Lett.} {\bf 114} (2015) 051301}, [\href{http://arxiv.org/abs/1406.4868}{{\tt
  1406.4868}}].

\bibitem{Bramante:2016rdh}
J.~Bramante, P.~J. Fox, G.~D. Kribs and A.~Martin, \emph{{Inelastic frontier:
  Discovering dark matter at high recoil energy}},
  \href{http://dx.doi.org/10.1103/PhysRevD.94.115026}{\emph{Phys. Rev.} {\bf
  D94} (2016) 115026}, [\href{http://arxiv.org/abs/1608.02662}{{\tt
  1608.02662}}].

\bibitem{Barello:2014uda}
G.~Barello, S.~Chang and C.~A. Newby, \emph{{A Model Independent Approach to
  Inelastic Dark Matter Scattering}},
  \href{http://dx.doi.org/10.1103/PhysRevD.90.094027}{\emph{Phys. Rev.} {\bf
  D90} (2014) 094027}, [\href{http://arxiv.org/abs/1409.0536}{{\tt
  1409.0536}}].

\bibitem{Gluscevic:2015sqa}
V.~Gluscevic, M.~I. Gresham, S.~D. McDermott, A.~H.~G. Peter and K.~M. Zurek,
  \emph{{Identifying the Theory of Dark Matter with Direct Detection}},
  \href{http://dx.doi.org/10.1088/1475-7516/2015/12/057}{\emph{JCAP} {\bf 1512}
  (2015) 057}, [\href{http://arxiv.org/abs/1506.04454}{{\tt 1506.04454}}].

\bibitem{Aprile:2017aas}
{\scshape XENON} collaboration, E.~Aprile et~al., \emph{{Effective field theory
  search for high-energy nuclear recoils using the XENON100 dark matter
  detector}}, \href{http://dx.doi.org/10.1103/PhysRevD.96.042004}{\emph{Phys.
  Rev.} {\bf D96} (2017) 042004}, [\href{http://arxiv.org/abs/1705.02614}{{\tt
  1705.02614}}].

\bibitem{Gelmini:2018ogy}
G.~B. Gelmini, V.~Takhistov and S.~J. Witte, \emph{{Casting a Wide Signal Net
  with Future Direct Dark Matter Detection Experiments}},
  \href{http://dx.doi.org/10.1088/1475-7516/2018/07/009}{\emph{JCAP} {\bf 1807}
  (2018) 009}, [\href{http://arxiv.org/abs/1804.01638}{{\tt 1804.01638}}].

\bibitem{Peter:2011eu}
A.~H.~G. Peter, \emph{{WIMP astronomy and particle physics with liquid-noble
  and cryogenic direct-detection experiments}},
  \href{http://dx.doi.org/10.1103/PhysRevD.83.125029}{\emph{Phys. Rev.} {\bf
  D83} (2011) 125029}, [\href{http://arxiv.org/abs/1103.5145}{{\tt
  1103.5145}}].

\bibitem{Drukier:1986tm}
A.~K. Drukier, K.~Freese and D.~N. Spergel, \emph{{Detecting Cold Dark Matter
  Candidates}}, \href{http://dx.doi.org/10.1103/PhysRevD.33.3495}{\emph{Phys.
  Rev.} {\bf D33} (1986) 3495--3508}.

\bibitem{Ho:2012bg}
C.~M. Ho and R.~J. Scherrer, \emph{{Anapole Dark Matter}},
  \href{http://dx.doi.org/10.1016/j.physletb.2013.04.039}{\emph{Phys. Lett.}
  {\bf B722} (2013) 341--346}, [\href{http://arxiv.org/abs/1211.0503}{{\tt
  1211.0503}}].

\bibitem{Gresham:2014vja}
M.~I. Gresham and K.~M. Zurek, \emph{{Effect of nuclear response functions in
  dark matter direct detection}},
  \href{http://dx.doi.org/10.1103/PhysRevD.89.123521}{\emph{Phys. Rev.} {\bf
  D89} (2014) 123521}, [\href{http://arxiv.org/abs/1401.3739}{{\tt
  1401.3739}}].

\bibitem{DelNobile:2018dfg}
E.~Del~Nobile, \emph{{A complete Lorentz-to-Galileo dictionary for direct Dark
  Matter detection}},  \href{http://arxiv.org/abs/1806.01291}{{\tt
  1806.01291}}.

\bibitem{Kavanagh:2018xeh}
B.~J. Kavanagh, P.~Panci and R.~Ziegler, \emph{{Faint Light from Dark Matter:
  Classifying and Constraining Dark Matter-Photon Effective Operators}},
  \href{http://arxiv.org/abs/1810.00033}{{\tt 1810.00033}}.

\bibitem{Mei:2005gm}
D.~Mei and A.~Hime, \emph{{Muon-induced background study for underground
  laboratories}},
  \href{http://dx.doi.org/10.1103/PhysRevD.73.053004}{\emph{Phys. Rev.} {\bf
  D73} (2006) 053004}, [\href{http://arxiv.org/abs/astro-ph/0512125}{{\tt
  astro-ph/0512125}}].

\bibitem{Akerib:2016mzi}
{\scshape LUX} collaboration, D.~S. Akerib et~al., \emph{{Low-energy (0.7-74
  keV) nuclear recoil calibration of the LUX dark matter experiment using D-D
  neutron scattering kinematics}},  \href{http://arxiv.org/abs/1608.05381}{{\tt
  1608.05381}}.

\bibitem{Polosatkin:2014dka}
S.~Polosatkin, E.~Grishnyaev and A.~Dolgov, \emph{{Liquid Argon Cryogenic
  Detector Calibration by Inelastic Scattering of Neutrons}},
  \href{http://arxiv.org/abs/1407.2718}{{\tt 1407.2718}}.

\bibitem{Akerib:2015cja}
{\scshape LZ} collaboration, D.~S. Akerib et~al., \emph{{LUX-ZEPLIN (LZ)
  Conceptual Design Report}},  \href{http://arxiv.org/abs/1509.02910}{{\tt
  1509.02910}}.

\bibitem{Aprile:2017aty}
{\scshape XENON} collaboration, E.~Aprile et~al., \emph{{The XENON1T Dark
  Matter Experiment}},
  \href{http://dx.doi.org/10.1140/epjc/s10052-017-5326-3}{\emph{Eur. Phys. J.}
  {\bf C77} (2017) 881}, [\href{http://arxiv.org/abs/1708.07051}{{\tt
  1708.07051}}].

\bibitem{Vietze:2014vsa}
L.~Vietze, P.~Klos, J.~Menéndez, W.~C. Haxton and A.~Schwenk, \emph{{Nuclear
  structure aspects of spin-independent WIMP scattering off xenon}},
  \href{http://dx.doi.org/10.1103/PhysRevD.91.043520}{\emph{Phys. Rev.} {\bf
  D91} (2015) 043520}, [\href{http://arxiv.org/abs/1412.6091}{{\tt
  1412.6091}}].

\bibitem{Bishara:2016hek}
F.~Bishara, J.~Brod, B.~Grinstein and J.~Zupan, \emph{{Chiral Effective Theory
  of Dark Matter Direct Detection}},
  \href{http://dx.doi.org/10.1088/1475-7516/2017/02/009}{\emph{JCAP} {\bf 1702}
  (2017) 009}, [\href{http://arxiv.org/abs/1611.00368}{{\tt 1611.00368}}].

\bibitem{Hoferichter:2015ipa}
M.~Hoferichter, P.~Klos and A.~Schwenk, \emph{{Chiral power counting of one-
  and two-body currents in direct detection of dark matter}},
  \href{http://dx.doi.org/10.1016/j.physletb.2015.05.041}{\emph{Phys. Lett.}
  {\bf B746} (2015) 410--416}, [\href{http://arxiv.org/abs/1503.04811}{{\tt
  1503.04811}}].

\bibitem{Hoferichter:2016nvd}
M.~Hoferichter, P.~Klos, J.~Menéndez and A.~Schwenk, \emph{{Analysis
  strategies for general spin-independent WIMP-nucleus scattering}},
  \href{http://dx.doi.org/10.1103/PhysRevD.94.063505}{\emph{Phys. Rev.} {\bf
  D94} (2016) 063505}, [\href{http://arxiv.org/abs/1605.08043}{{\tt
  1605.08043}}].

\bibitem{Green:2007rb}
A.~M. Green, \emph{{Determining the WIMP mass using direct detection
  experiments}},
  \href{http://dx.doi.org/10.1088/1475-7516/2007/08/022}{\emph{JCAP} {\bf 0708}
  (2007) 022}, [\href{http://arxiv.org/abs/hep-ph/0703217}{{\tt
  hep-ph/0703217}}].

\bibitem{Green:2008rd}
A.~M. Green, \emph{{Determining the WIMP mass from a single direct detection
  experiment, a more detailed study}},
  \href{http://dx.doi.org/10.1088/1475-7516/2008/07/005}{\emph{JCAP} {\bf 0807}
  (2008) 005}, [\href{http://arxiv.org/abs/0805.1704}{{\tt 0805.1704}}].

\bibitem{McDermott:2011hx}
S.~D. McDermott, H.-B. Yu and K.~M. Zurek, \emph{{The Dark Matter Inverse
  Problem: Extracting Particle Physics from Scattering Events}},
  \href{http://dx.doi.org/10.1103/PhysRevD.85.123507}{\emph{Phys. Rev.} {\bf
  D85} (2012) 123507}, [\href{http://arxiv.org/abs/1110.4281}{{\tt
  1110.4281}}].

\bibitem{Feroz:2008xx}
F.~Feroz, M.~P. Hobson and M.~Bridges, \emph{{MultiNest: an efficient and
  robust Bayesian inference tool for cosmology and particle physics}},
  \href{http://dx.doi.org/10.1111/j.1365-2966.2009.14548.x}{\emph{Mon. Not.
  Roy. Astron. Soc.} {\bf 398} (2009) 1601--1614},
  [\href{http://arxiv.org/abs/0809.3437}{{\tt 0809.3437}}].

\bibitem{Fowlie:2016hew}
A.~Fowlie and M.~H. Bardsley, \emph{{Superplot: a graphical interface for
  plotting and analysing MultiNest output}},
  \href{http://dx.doi.org/10.1140/epjp/i2016-16391-0}{\emph{Eur. Phys. J. Plus}
  {\bf 131} (2016) 391}, [\href{http://arxiv.org/abs/1603.00555}{{\tt
  1603.00555}}].

\bibitem{Cerdeno:2018bty}
D.~G. Cerde\~no, A.~Cheek, E.~Reid and H.~Schulz, \emph{{Surrogate Models for
  Direct Dark Matter Detection}},
  \href{http://dx.doi.org/10.1088/1475-7516/2018/08/011}{\emph{JCAP} {\bf 1808}
  (2018) 011}, [\href{http://arxiv.org/abs/1802.03174}{{\tt 1802.03174}}].

\bibitem{Bozorgnia:2016ogo}
N.~Bozorgnia, F.~Calore, M.~Schaller, M.~Lovell, G.~Bertone, C.~S. Frenk
  et~al., \emph{{Simulated Milky Way analogues: implications for dark matter
  direct searches}},
  \href{http://dx.doi.org/10.1088/1475-7516/2016/05/024}{\emph{JCAP} {\bf 1605}
  (2016) 024}, [\href{http://arxiv.org/abs/1601.04707}{{\tt 1601.04707}}].

\bibitem{Kelso:2016qqj}
C.~Kelso, C.~Savage, M.~Valluri, K.~Freese, G.~S. Stinson and J.~Bailin,
  \emph{{The impact of baryons on the direct detection of dark matter}},
  \href{http://dx.doi.org/10.1088/1475-7516/2016/08/071}{\emph{JCAP} {\bf 1608}
  (2016) 071}, [\href{http://arxiv.org/abs/1601.04725}{{\tt 1601.04725}}].

\bibitem{Sloane:2016kyi}
J.~D. Sloane, M.~R. Buckley, A.~M. Brooks and F.~Governato, \emph{{Assessing
  Astrophysical Uncertainties in Direct Detection with Galaxy Simulations}},
  \href{http://dx.doi.org/10.3847/0004-637X/831/1/93}{\emph{Astrophys. J.} {\bf
  831} (2016) 93}, [\href{http://arxiv.org/abs/1601.05402}{{\tt 1601.05402}}].

\bibitem{Bozorgnia:2017brl}
N.~Bozorgnia and G.~Bertone, \emph{{Implications of hydrodynamical simulations
  for the interpretation of direct dark matter searches}},
  \href{http://dx.doi.org/10.1142/S0217751X17300162}{\emph{Int. J. Mod. Phys.}
  {\bf A32} (2017) 1730016}, [\href{http://arxiv.org/abs/1705.05853}{{\tt
  1705.05853}}].

\bibitem{Bertone:2007xj}
G.~Bertone, D.~G. Cerde\~no, J.~I. Collar and B.~C. Odom, \emph{{WIMP
  identification through a combined measurement of axial and scalar
  couplings}},
  \href{http://dx.doi.org/10.1103/PhysRevLett.99.151301}{\emph{Phys. Rev.
  Lett.} {\bf 99} (2007) 151301}, [\href{http://arxiv.org/abs/0705.2502}{{\tt
  0705.2502}}].

\bibitem{Pato:2010zk}
M.~Pato, L.~Baudis, G.~Bertone, R.~Ruiz~de Austri, L.~E. Strigari and
  R.~Trotta, \emph{{Complementarity of Dark Matter Direct Detection Targets}},
  \href{http://dx.doi.org/10.1103/PhysRevD.83.083505}{\emph{Phys. Rev.} {\bf
  D83} (2011) 083505}, [\href{http://arxiv.org/abs/1012.3458}{{\tt
  1012.3458}}].

\bibitem{Cerdeno:2013gqa}
D.~G. Cerdeño et~al., \emph{{Complementarity of dark matter direct detection:
  the role of bolometric targets}},
  \href{http://dx.doi.org/10.1088/1475-7516/2013/07/028,
  10.1088/1475-7516/2013/09/E01}{\emph{JCAP} {\bf 1307} (2013) 028},
  [\href{http://arxiv.org/abs/1304.1758}{{\tt 1304.1758}}].

\bibitem{Peter:2013aha}
A.~H.~G. Peter, V.~Gluscevic, A.~M. Green, B.~J. Kavanagh and S.~K. Lee,
  \emph{{WIMP physics with ensembles of direct-detection experiments}},
  \href{http://dx.doi.org/10.1016/j.dark.2014.10.006}{\emph{Phys. Dark Univ.}
  {\bf 5-6} (2014) 45--74}, [\href{http://arxiv.org/abs/1310.7039}{{\tt
  1310.7039}}].

\bibitem{Catena:2014uqa}
R.~Catena and P.~Gondolo, \emph{{Global fits of the dark matter-nucleon
  effective interactions}},
  \href{http://dx.doi.org/10.1088/1475-7516/2014/09/045}{\emph{JCAP} {\bf 1409}
  (2014) 045}, [\href{http://arxiv.org/abs/1405.2637}{{\tt 1405.2637}}].

\bibitem{Edwards:2018lsl}
T.~D.~P. Edwards, B.~J. Kavanagh and C.~Weniger, \emph{{Dark Matter Model or
  Mass, but Not Both: Assessing Near-Future Direct Searches with Benchmark-free
  Forecasting}},  \href{http://arxiv.org/abs/1805.04117}{{\tt 1805.04117}}.

\end{thebibliography}\endgroup

\end{document}